\documentclass[11pt,twoside]{article}

\usepackage{asp2006}
\usepackage{amsmath}
\usepackage{txfonts}
\usepackage{graphicx}
\begin{document}
\title{The Richness and Beauty of the Physics of Cosmological Recombination}
\author{R.A. Sunyaev\altaffilmark{1,2} and J. Chluba\altaffilmark{1}}

%%% Fill in author affiliations
\affil{$^1$Max-Planck-Institut f\"ur Astrophysik, Karl-Schwarzschild-Str. 1,
85741 Garching bei M\"unchen, Germany
\\
$^2$Space Research Institute, Russian Academy of Sciences, Profsoyuznaya 84/32,
117997 Moscow, Russia
}    

%%% Fill in author affiliations
%\altaffiltext{1}{Max-Planck-Institut f\"ur Astrophysik, Karl-Schwarzschild-Str. 1,
%85741 Garching bei M\"unchen, Germany}

%\altaffiltext{2}{Space Research Institute, Russian Academy of Sciences, Profsoyuznaya 84/32,
%117997 Moscow, Russia}    

\begin{abstract} %%% Abstract to run on from here.
  The physical ingredients to describe the epoch of cosmological recombination
  are amazingly simple and well-understood. This fact allows us to take into
  account a very large variety of processes, still finding potentially
  measurable consequences.
  In this contribution we highlight some of the detailed physics that were
  recently studied in connection with cosmological hydrogen recombination. The
  impact of these considerations is two-fold: 
  (i) the associated release of photons during this epoch leads to interesting
  and {\it unique deviations} of the Cosmic Microwave Background (CMB) energy
  spectrum {\it from a perfect blackbody}, which, in particular at decimeter
  wavelength, may become observable in the near future.
  Observing these distortions, in principle would provide an additional way to
  determine some of the key parameters of the Universe (e.g. the specific
  entropy, the CMB monopole temperature and the pre-stellar abundance of
  helium), {\it not suffering} from limitations set by {\it cosmic variance}. 
  Also it permits us to confront our detailed understanding of the
  recombination process with {\it direct observational evidence}.
  In this contribution we illustrate how the theoretical {\it spectral template} for the
  cosmological recombination spectrum may be utilized for this purpose.
  (ii) with the advent of high precision CMB data, e.g. as will be available
  using the {\sc Planck} Surveyor or {\sc Cmbpol}, a very accurate theoretical
  understanding of the {\it ionization history} of the Universe becomes
  necessary for the interpretation of the CMB temperature and polarization
  anisotropies.
  Here we show that the uncertainty in the ionization history due to several
  processes that until now are not taken in to account in the standard
  recombination code {\sc Recfast} exceed the level of $0.1\%$ to $0.5\%$ for
  each of them.
  However, it is indeed surprising how {\it inert} the cosmological
  recombination history is even at percent-level accuracy.

\end{abstract}

%%% MAIN BODY OF TEXT GOES HERE. CONSULT "INSTRUCTIONS FOR AUTHORS USING
%%% LATEX2E MARKUP", SECTIONS 2.3-2.6 FOR HELP WITH EQUATIONS, FIGURES,
%%% AND TABLES.

\section{Introduction.}
\label{RS:sec:Intro}
%---------------

\subsection{What is so rich and beautiful about cosmological recombination?}
\label{RS:sec:Intro1}
%---------------
Within the cosmological concordance model the physical environment during the
epoch of cosmological recombination (redshifts $500 \lesssim z\lesssim 2000$
for hydrogen, $1600 \lesssim z\lesssim 3500$ for
\ion{He}{II}$\rightarrow$\ion{He}{I} and $5000 \lesssim z\lesssim 8000$ for
\ion{He}{III}$\rightarrow$\ion{He}{II} recombination) is extremely simple: the
Universe is homogeneous and isotropic, globally neutral and is expanding at a
rate that can be computed knowing a small set of cosmological parameters.
%
%%, at those times are mainly the matter energy density and to a smaller extend
%%the energy density due to relativistic particles (photons and neutrinos).
%
The baryonic matter component is dominated by hydrogen ($\sim 76\%$) and
helium ($\sim 24\%$), with negligibly small traces of other light
elements, such as deuterium and lithium, and it is continuously exposed to a
bath of isotropic blackbody radiation, which contains roughly $1.6\times 10^9$
photons per baryon.
These initially simple and very unique settings in principle allows us to
predict the {\it ionization history} of the Universe and the {\it
  cosmological recombination spectrum} (see Sect.~\ref{RS:sec:spectrum}) with
extremely high accuracy, where the limitations are mainly set by our
understanding of the {\it atomic processes} and associated transition rates.
It is this simplicity that offers us the possibility to enter a {\it
  rich} field of physical processes and to challenge our understanding of
atomic physics, radiative transfer and cosmology, eventually leading to a {\it
  beautiful} variety of potentially observable effects.

\subsection{What is so special about cosmological recombination?}
\label{RS:sec:Intro2}
%---------------
%
The main reason for this simplicity is the {\it extremely large specific
entropy} and the {\it slow expansion} of our Universe.
Due to the huge number of CMB photons, the free electrons are tightly coupled
to the radiation field due to tiny energy exchange during {\it Compton
    scattering} off thermal electrons until rather low redshifts, such that
during recombination the thermodynamic temperature of electrons is equal to
the CMB blackbody temperature with very high precision.
In addition, the very fast {\it Coulomb interaction} and atom-ion
  collisions allows to maintain full thermodynamic equilibrium among the
electrons, ions and neutral atoms down to $z\sim 150$ \citep{RS_Zeldovich68}.
Also, processes in the baryonic sector cannot severely affect any of the
radiation properties, down to redshift where the first stars and galaxies
appear,
and the atomic rates are largely dominated by radiative processes, including
{\it stimulated recombination}, {\it induced emission} and absorption of
photons.
On the other hand, the slow expansion of the Universe allows us to consider
the evolution of the atomic species along a sequence of {\it quasi-stationary}
stages, where the populations of the levels are nearly in full equilibrium
with the radiation field, but only subsequently and very slowly drop out of
equilibrium, finally leading to {\it recombination} and the {\it release of
  additional photons} in uncompensated bound-bound and free-bound transitions.

\subsection{Historical overview.}
\label{RS:sec:history}
%---------------
%The recombination of hydrogen, mediated by the expansion of the
%Universe, was predicted by {\it George Gamow}, in one of his first papers
%about the Big Bang model of the Universe \citep{RS_Gamow}.
%
It was realized at the end of the 60's \citep{RS_Zeldovich68, RS_Peebles68},
that during the epoch of cosmological hydrogen recombination (typical
redshifts $800\lesssim z \lesssim 1600$) any direct recombination of electrons
to the ground state of hydrogen is immediately followed by the ionization of a
neighboring neutral atom due to re-absorption of the newly released
Lyman-continuum photon.
In addition, because of the enormous difference in the $2{\rm
  p}\leftrightarrow 1{\rm s}$ dipole transition rate and the Hubble expansion
{rate}, photons emitted close to the center of the Lyman-$\alpha$
line scatter $\sim 10^8-10^9$ times before they can finally escape further
interaction with the medium and thereby permit a successful settling
  of electrons in the 1s-level.
  It is due to these very peculiar circumstances that the $2{\rm
    s}\leftrightarrow 1{\rm s}$-two-photon decay process, being $\sim 10^8$
  orders of magnitude slower than the Lyman-$\alpha$ resonance transition, is
  able to substantially control the dynamics of cosmological hydrogen
  recombination \citep{RS_Zeldovich68, RS_Peebles68},
  allowing about 57\% of all hydrogen atoms in the Universe to recombine at
  redshift $z\lesssim 1400$ through this channel \citep{RS_Chluba2006b}.

% Anisotropies
Shortly afterwards \citep{RS_Sunyaev1970, RS_Peebles1970}, the
importance of the ionization history as one of the key ingredients for the
theoretical predictions of the Cosmic Microwave Background (CMB) temperature
and polarization anisotropies became clear, and today these tiny directional
variations of the CMB temperature ($\Delta T/T_0\sim 10^{-5}$) around the mean
value $T_0=2.725\pm 0.001\,$K \citep{RS_Fixsen2002} have been
observed for the whole sky using the 
{\sc
  Cobe}        %\footnote{http://lambda.gsfc.nasa.gov/product/cobe/} 
and {\sc Wmap} %\footnote{http://lambda.gsfc.nasa.gov/product/map/current/}
satellites, beyond doubt with great success.
The high quality data coming from balloon-borne and ground-based CMB
experiments ({\sc Boomerang, Maxima, Archeops, Cbi, Dasi} and {\sc Vsa} etc.)
today certainly provides one of the mayor pillars for the {\it cosmological
concordance model} \citep{RS_Bennett2003}, and planned CMB mission like the
{\sc Planck} Surveyor 
%
%\footnote{www.rssd.esa.int/Planck} 
%
will help to further establish the {\it era of precision cosmology}.

%
%---------------
\begin{figure}
\centering 
\includegraphics[width=0.8\columnwidth]{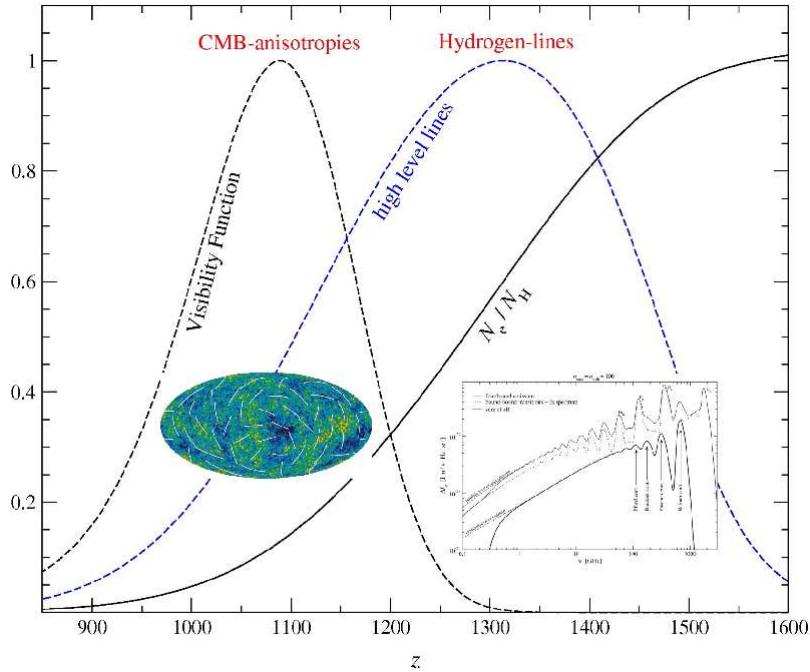}
\caption{Ionization history of the Universe and the origin of CMB signals. 
  The observed angular fluctuations in the CMB temperature are created close
  to the maximum of the Thomson visibility function around $z\sim 1089$,
  whereas the direct information carried by the photons in the cosmological
  hydrogen recombination spectrum is from earlier times.}
\label{RS:fig:plot1}
\end{figure}
%-------------------------------------
%=====================================
% spectrum
%=====================================
In September 1966, one of the authors (RS) was explaining during a seminar at
  the Shternberg Institute in Moscow how according to the Saha formula this
  recombination should occur.
  After the talk his friend (UV astronomer) {\it Vladimir Kurt}
  asked him: {\it 'but where are all the redshifted Lyman-$\alpha$ photons
    that were released during recombination?'}
  Indeed this was a great question, which was then addressed in detail
  by \citet{RS_Zeldovich68}, leading to an understanding of the role of the
  2s-two-photon decay, the delay of recombination as compared to the
  Saha-solution, the spectral distortions of the CMB due to two-photon
  continuum and Lyman-$\alpha$ emission, the frozen remnant of ionized atoms,
  and the radiation and matter temperature equality until $z\sim 150$.

%Also in the late 60's it was realized that there should be some additional
%amount of radiation connected with the transition of the full-ionized medium
%to the neutral state.
%%
%After a talk of one of the authors about equilibrium recombination in 1968,
%{\it Vladimir Kurt}, an ....., was asking: 'I do not understand. Where are all
%the Lyman-$\alpha$ photons from recombination?'. Indeed this was a great
%question, which was then addressed a bit later by \citet{RS_Zeldovich68}.

%
All recombined electrons in hydrogen lead to the release of $\sim
13.6\,$eV in form of photons, but due to the large specific entropy of the Universe
%(there are $\sim 1.6\times 10^9$ photons per baryon) 
this will only add some fraction of $\Delta \rho_\gamma/\rho_\gamma\sim
10^{-9}-10^{-8}$ to the total energy density of the CMB spectrum, and hence
the corresponding distortions are expected to be very small.
However, all the photons connected with the Lyman-$\alpha$ transition and the
2s-two-photon continuum today appear in the Wien part of the CMB spectrum,
where the number of photons in the CMB blackbody is dropping exponentially,
and, as realized earlier \citep{RS_Zeldovich68, RS_Peebles68}, these
distortions are significant (see Sect.~\ref{RS:sec:spectrum}).

In 1975, {\it Victor Dubrovich} pointed out that the transitions among highly
excited levels in hydrogen are producing additional photons, which after
redshifting are reaching us in the cm-spectral band. This band is actually
accessible from the ground.
%  , but there the amplitude of the considered distortions is much smaller than
%  those in the Wien part.
%
Later these early estimates were significantly refined by several groups
(e.g. see \citet{RS_Kholu2005} and \citet{RS_Jose2006} for references),
with the most recent calculation performed by \citet{RS_Chluba2006b}, also
including the previously neglected free-bound component, and showing in detail
that the relative distortions are becoming more significant in the
decimeter Rayleigh-Jeans part of the CMB blackbody spectrum
(Fig.~\ref{RS:fig:DI_results}).
These kind of precise computations are becoming feasible today, because (i)
our knowledge of atomic data (in particular for neutral helium) has
significantly improved, and (ii) it is now possible to handle large systems of
strongly coupled differential equations using modern computers.
The most interesting aspect of this radiation is that it has a very {\it
  peculiar} but {\it well-defined, quasi-periodic spectral dependence}, where
the photons are coming from redshifts $z\sim 1300-1400$, i.e. {\it before} the
time of the formation of the CMB angular fluctuation close to the maximum of
the Thomson visibility function (see Fig.~\ref{RS:fig:plot1}).
Therefore, measuring these distortions of the CMB spectrum would provide a way
to confront our understanding of the recombination epoch with {\it direct
  experimental evidence},
and in principle may open another independent way to determine some of the key
parameters of the Universe, in particular the value of the CMB monopole
temperature, $T_0$, the number density of baryons, $\propto \Omega_{\rm
  b}h^2$, or alternatively the specific entropy, and the primordial helium
abundance (e.g.  see \citet{RS_Chluba2007d} and references therein).

%-------------------------------------
\section{The hydrogen recombination spectrum.}
\label{RS:sec:spectrum}
%---------------
%---------------
\begin{figure}
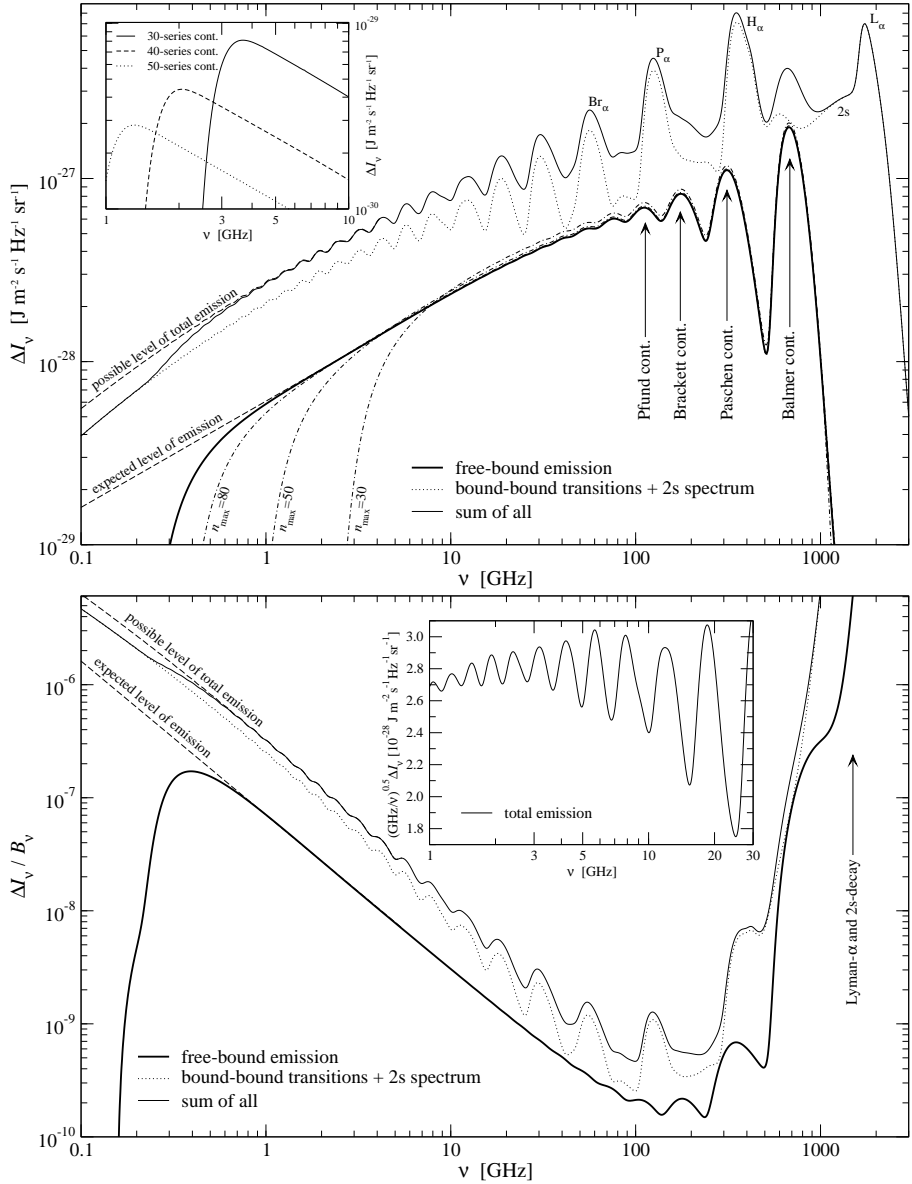

\centering 
\includegraphics[width=0.9\columnwidth]{./RS.DI.eps}
\\
\includegraphics[width=0.9\columnwidth]{./RS.DI_I.eps}
\caption{The full hydrogen recombination spectrum including the free-bound
  emission.
  The results of the computation for 100 shells were used.
%  as presented in \citet{RS_Chluba2007}.
%
The contribution due to the 2s two-photon decay is also accounted for.
The dashed lines indicate the expected level of emission when including more
shells. In the upper panel we also show the free-bound continuum spectrum for
different values of $n_{\rm max}$ (dashed-dotted). The inlay gives the
free-bound emission for $n=30,\,40$, and $50$.
The lower panel shows the distortion relative to the CMB blackbody spectrum,
and the inlay illustrates the modulation of the total emission spectrum for
$1\,\text{GHz}\leq \nu \leq 30\,\text{GHz}$ in convenient coordinates. The
figure is from \citet{RS_Chluba2006b}.}
\label{RS:fig:DI_results}
\end{figure}
%-------------------------------------
Within the picture described above it is possible to compute the {\it
  cosmological hydrogen recombination spectrum} with high accuracy.
In Figure \ref{RS:fig:DI_results} we give the results of our computations for
frequencies from $100\,$MHz up to $3000\,$GHz.
The free-bound and bound-bound atomic {\it transitions among 5050 atomic
  levels} had to be taken into account in these computations.
At high frequencies one can clearly see the features connected with the
Lyman-$\alpha$ line, and the Balmer-, Paschen- and Brackett-series, whereas
below $\nu\sim 1\,$GHz the lines coming from transitions between highly
excited level start to merge to a continuum.
Also the features due to the Balmer and the 2s-1s two-photon continuum are
visible.
%
%Overall the free-bound emission contributes about 20-30\% to the spectral
%distortion due to hydrogen recombination at each frequency, and a total of
%$\sim 5$ photons per hydrogen atom are released in the full hydrogen
%recombination spectrum.
%
In total $\sim 5$ photons per hydrogen atom are released in the full hydrogen
recombination spectrum.

%%
%The most interesting aspect of this radiation is that it has a very {\it
%  peculiar} but {\it well-defined spectral dependence}, where the photons are
%coming from redshifts $z\sim 1300-1400$, so {\it before} the time of the
%formation of the CMB angular fluctuation close to the maximum of the Thomson
%visibility function. Therefore, measuring these distortions of the CMB
%spectrum will provide a way to confront our understanding of the recombination
%epoch with direct experimental evidence.

%
One can also see from Figure \ref{RS:fig:DI_results} that both in the Wien and
the Rayleigh-Jeans region of the CMB blackbody spectrum the relative
distortion is growing. In the vicinity of the Lyman-$\alpha$ line the relative
distortion exceed unity by several orders of magnitude, but unfortunately at
these frequencies the cosmic infra-red background due to submillimeter, dusty
galaxies renders a direct measurement impossible.
Similarly, around the maximum of the CMB blackbody at $\sim 150\,$GHz it will
be hard to measure these distortions with current technology, although there
the spectral variability of the recombination radiation is largest.  
However, at low frequencies the relative distortion exceeds the level of
$\Delta I/I\sim 10^{-7}$ at frequency $\nu\sim 2\,$GHz but still has
variability with well-defined frequency dependence at a level of several
percent.

%---------------
\begin{figure}
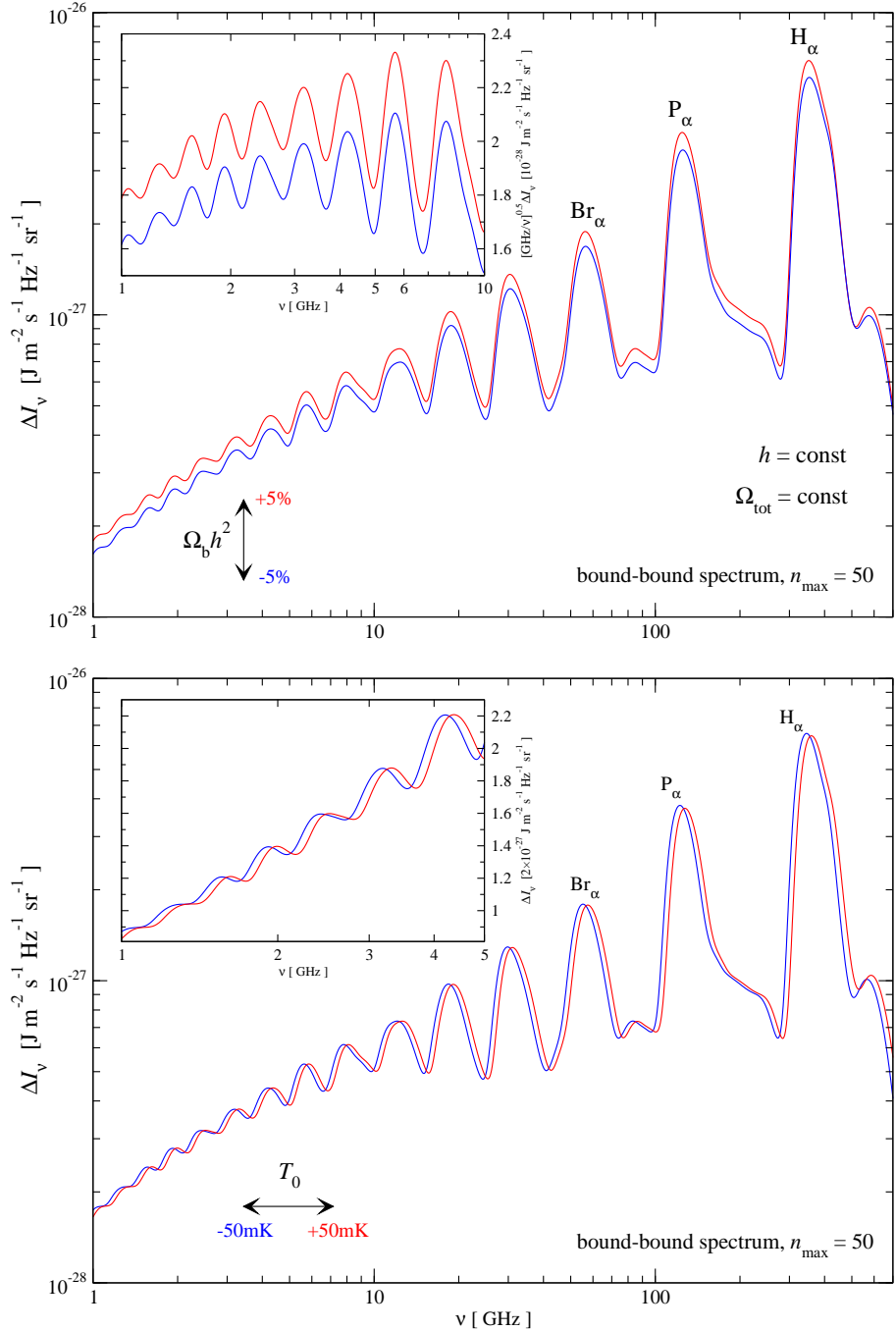

\centering 
%\includegraphics[width=0.9\columnwidth]{./RS:DJ.h.Obh2.50.eps}
%\\
\includegraphics[width=0.9\columnwidth]{./RS.DJ.Ob.50.eps}
\\
\includegraphics[width=0.9\columnwidth]{./RS.DJ.T0.50mK.50.eps}
\caption{The bound-bound hydrogen recombination spectrum for $n_{\rm
    max}=50$. The upper panel illustrates the dependence on $\Omega_{\rm
    b}h^2$, and the lower the dependence on the value of $T_0$. 
The figure is from \citet{RS_Chluba2007c}.}
\label{RS:fig:DI_cosmos}
\end{figure}
%-------------------------------------
\subsection{Dependence the recombination spectrum on cosmological parameters.}
\label{RS:sec:spectrum_cosmos}
%---------------
%---------------
In this Section we want to {\it illustrate} the impact of different
cosmological parameters on the hydrogen recombination spectrum. We restrict
ourselves to the bound-bound emission spectrum and included only 50 shells for
the hydrogen atom into our computations. A more rigorous investigation is in
preparation, however, the principle conclusions should not be affected.

In Fig.~\ref{RS:fig:DI_cosmos} we illustrate the dependence of the hydrogen
recombination spectrum on the value of the CMB monopole temperature, $T_0$.
The value of $T_0$ mainly defines the time of recombination, and consequently
when most of the emission in each transition appears. This leads to a
dependence of the line positions on $T_0$, but the total intensity in each
transition (especially at frequencies $\nu \lesssim 30\,$GHz) remains
practically the same. We found that the fractional shift of the low frequency
spectral features along the frequency axis scales roughly like $\Delta
\nu/\nu\sim\Delta T/T_0$. 
%
%Hence $\Delta T\sim1\,$mK implies $\Delta\nu/\nu\sim 0.04\,\%$ or
%$\Delta\nu\sim 1\,$MHz at $2\,$GHz.
%
Since the maxima and minima of the line features due to the large duration of
recombination are rather broad ($\sim 10-20\%$), it is probably better to look
for these shifts close to the steep parts of the lines, where the derivatives
of the spectral distortion due to hydrogen recombination are largest.
It is also important to mention that the hydrogen recombination spectrum is
shifted as a {\it whole}, allowing to increase the significance of a
measurement by considering many spectral features at several frequencies.

We showed in \citet{RS_Chluba2007c} that the cosmological hydrogen
recombination spectrum is practically independent of the value of $h$. Only
the features due to the Lyman, Balmer, Paschen and Brackett series are
slightly modified.
This is connected to the fact, that $h$ affects the ratio of the atomic
time-scales to the expansion time. Therefore changing $h$ affects the escape
rate of photons in the Lyman-$\alpha$ transition and the relative importance
of the 2s-1s transition. For transitions among highly excited states it is not
crucial via which channel the electrons finally reach the ground state of
hydrogen and hence the modifications of the recombination spectrum at low
frequencies due to changes of $h$ are small.
Changes of $\Omega_{\rm m}h^2$ should affect the recombination spectrum for
the same reason.

The lower panel in Fig.~\ref{RS:fig:DI_cosmos} illustrates the dependence of
the hydrogen recombination spectrum on $\Omega_{\rm b}h^2$. It was shown that
the total number of photons released during hydrogen recombination is directly
related to the total number of hydrogen nuclei \citep[e.g.][]{RS_Chluba2007d}.
Therefore one expects that the overall normalization of the recombination
spectrum depends on the total number of baryons, $N_{\rm b}\propto\Omega_{\rm
  b}h^2$, and the helium to hydrogen abundance ratio, $Y_{\rm p}$.
Varying $\Omega_{\rm b}h^2$ indeed leads to a change in the overall amplitude
$\propto \Delta(\Omega_{\rm b}h^2)/(\Omega_{\rm b}h^2)$.
Similarly, changes of $Y_{\rm p}$ should affect the normalization of the
hydrogen recombination spectrum, but here it is important to also take the
helium recombination spectrum into account. 
%
%Like in the case of hydrogen there should be an effective number of photons
%that is produced per helium atom during \ion{He}{III}$\rightarrow$\ion{He}{II}
%and \ion{He}{II}$\rightarrow$\ion{He}{I} recombination. Changing $Y_{\rm p}$
%will affect the relative contribution of the helium to the hydrogen
%recombination spectrum. Since the physics of helium recombination is different
%than in the case of hydrogen (e.g. the spectrum of neutral helium is more
%complicated; helium recombination occurs at earlier times, when the medium was
%hotter; \ion{He}{III}$\rightarrow$\ion{He}{II} is more rapid, so that the
%recombination lines are more narrow \citep{RS_Chluba2007d}, one can expect to
%find direct evidence of the presence of helium in the full recombination
%spectrum. These might be used to quantify the total amount of helium during
%the epoch of recombination, well before the first appearance of stars.

%---------------
\begin{figure}
\centering 
\includegraphics[width=0.8\columnwidth]{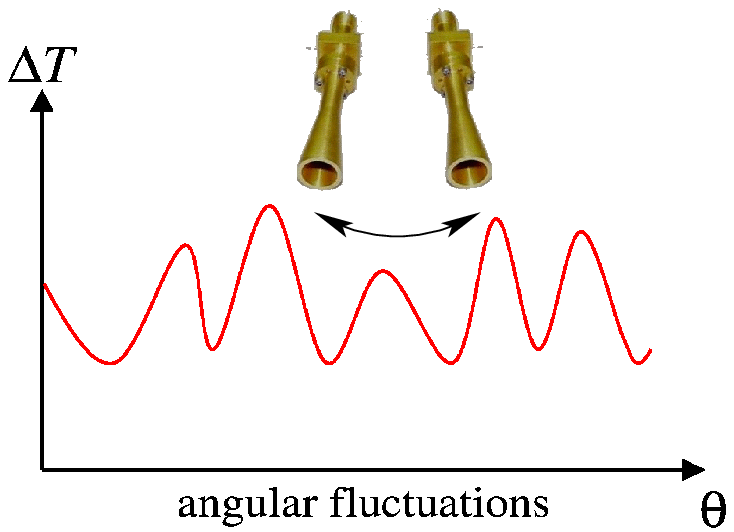}
\\[8mm]
\includegraphics[width=0.8\columnwidth]{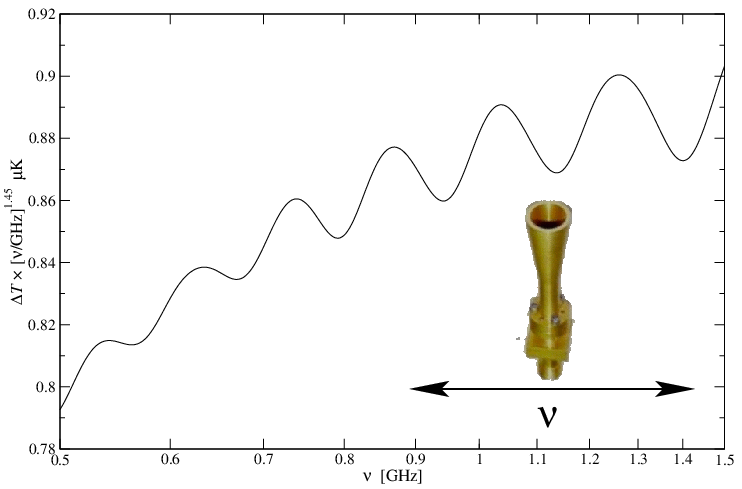}
\caption{Comparison of observing strategies: top panel -- observations of the
  CMB angular fluctuations. Here one is scanning the sky at fixed frequency in
  different directions. lower panel -- proposed strategy for the signal from
  cosmological recombination. For this one may fix the observing direction,
  choosing a large, least contaminated part of the sky, and scan along the
  frequency axis instead.}
\label{RS:fig:strategy}
\end{figure}
%-------------------------------------
%-------------------------------------
\subsection{A possible observing strategy.}
\label{RS:sec:spectrum_strat}
%---------------
In order to measure these distortions one should scan the CMB spectrum along
the {\it frequency axis} including several spectral bands (for illustration
see Fig.~\ref{RS:fig:strategy}). Because the CMB spectrum is the same in all
directions, one can collect the flux of large regions on the sky, particularly
choosing patches that a the least contaminated by other astrophysical
foregrounds.
{\it No absolute measurement} is necessary, but one only has to look for a
modulated signal at the $\sim\mu$K level, with typical amplitude of
$\sim 30\,$nK and $\Delta \nu/\nu\sim 0.1$, where this signal can be predicted
with high accuracy, yielding a {\it spectral template} for the full
cosmological recombination spectrum, which should also include the
contributions from helium.
For observations of the CMB angular fluctuations a sensitivity level of
$10\,$nK in principle can be already achieved \citep{RS_ReadheadPC}.

%Here we want to stress again, that measuring these distortions of the CMB
%spectrum would provide a way to confront our understanding of the
%recombination epoch with {\it direct experimental evidence},
%%
%and in principle may open another independent way to determine some of the key
%parameters of the Universe, in particular the value of the CMB monopole
%temperature, $T_0$, the number density of baryons, $\propto \Omega_{\rm
%  b}h^2$, or alternatively the specific entropy, and the pre-stellar helium
%abundance, {\it not suffering} from limitations set by {\it cosmic
%    variance}.
%

\section{Previously neglected physical processes during hydrogen recombination.}
\label{RS:sec:processes}
%---------------
With the improvement of available CMB data also refinements of the
computations regarding the ionization history became necessary, leading to the
development of the widely used {\sc Recfast} code \citep{RS_Seager1999,
  RS_Seager2000}.
However, the prospects with the {\sc Planck} Surveyor have motivated several
groups to re-examine the problem of cosmological hydrogen and helium
recombination,
with the aim to identify previously neglected physical processes that could
affect the ionization history of the Universe at the level of $\gtrsim 0.1\%$.
Such accuracy becomes necessary to achieve the promised precision for the
estimation of cosmological parameters using the observation of the CMB angular
fluctuations and acoustic peaks.
Here we wish to provide an overview of the most important additions in this
context and to highlight some of the previously neglected physical processes
in particular.
Most of them are also important during the epoch of helium recombination
\citep[e.g.][]{RS_HirataI}, but here we focus our discussion on hydrogen only.
The superposition of all effects listed below lead to an ambiguity in
the ionization history during the epoch cosmological hydrogen recombination
that exceeds the level of 0.1\% to 0.5\% each, and therefore should
be taken into account in the detailed analysis of future CMB data.
Still this shows that the simple picture, as explained in
Sect.~\ref{RS:sec:Intro} is amazingly stable.

%%---------------
%\begin{figure}
%\centering 
%\includegraphics[width=0.8\columnwidth]{./RS:Rec_coeffi_lmax.corr.eps}
%%
%\caption{$l$-dependence of the recombination coefficient, $\alpha_{nl}$, at
%$z=1300$ for different shells. The curves have been re-scaled by the {\it
%total} recombination coefficient, $\alpha_{n,\rm tot}=\sum_l \alpha_{nl}$, and
%multiplied by $l_{\rm max}=n-1$ such that the 'integral' over $\xi=l/l_{\rm
%max}$ becomes unity. Also the results obtained within the Kramers'
%approximation, i.e. $\alpha_{nl}^{\rm K}={\rm const}\times[2l+1]$, and without
%the inclusion of stimulated recombination for $n=100$ are presented. The
%figure is from \citet{RS_Chluba2007}.}
%\label{RS:fig:Rci_Rci_tot}
%\end{figure}
%%---------------
\subsubsection{Detailed evolution of the populations in the angular momentum sub-states.}
\label{RS:sec:pops}
%---------------
The numerical solution of the hydrogen recombination history and the
associated spectral distortions of the CMB requires the integration of a stiff
system of coupled ordinary differential equations, describing the evolution of
the populations of the different hydrogen levels, with extremely high
accuracy.
Until now this task was only completed using additional simplifying
assumptions.
Among these the most important simplification is to assume {\it full
  statistical equilibrium}.
%\footnote{i.e. the population of a given level $(n,l)$ is determined by
%  $N_{nl} = (2l+1) N_n / n^2$, where $N_n$ is the total population of the
%  shell with principle quantum number $n$.}  (SE) within a given shell for
%$n>2$.
%(for a more detailed comparison of the different approached see
%\citet{RS_Jose2006} and references therein).
%
However, as was shown in \citet{RS_Jose2006} and \citet{RS_Chluba2007}, this
leads to an overestimation of the hydrogen recombination rate at low redshift
by up to $\sim10\%$. This is mainly because during hydrogen recombination
collision are so much weaker than radiative processes, so that the populations
within a given atomic shell depart from SE. It was also shown that for the
highly excited level stimulated emission and recombination (see
Fig.~\ref{RS:fig:Rci_Rci_tot}) is important.
%has to be taken into account.

\subsubsection{Induced two-photon decay of the hydrogen 2s-level.}
\label{RS:sec:2gamma2s}
%---------------
%-----------------------------
\begin{figure}
  \centering 
  \includegraphics[width=0.8\columnwidth]{./RS.Rec_coeffi_lmax.corr.eps}
  \includegraphics[width=0.8\columnwidth]{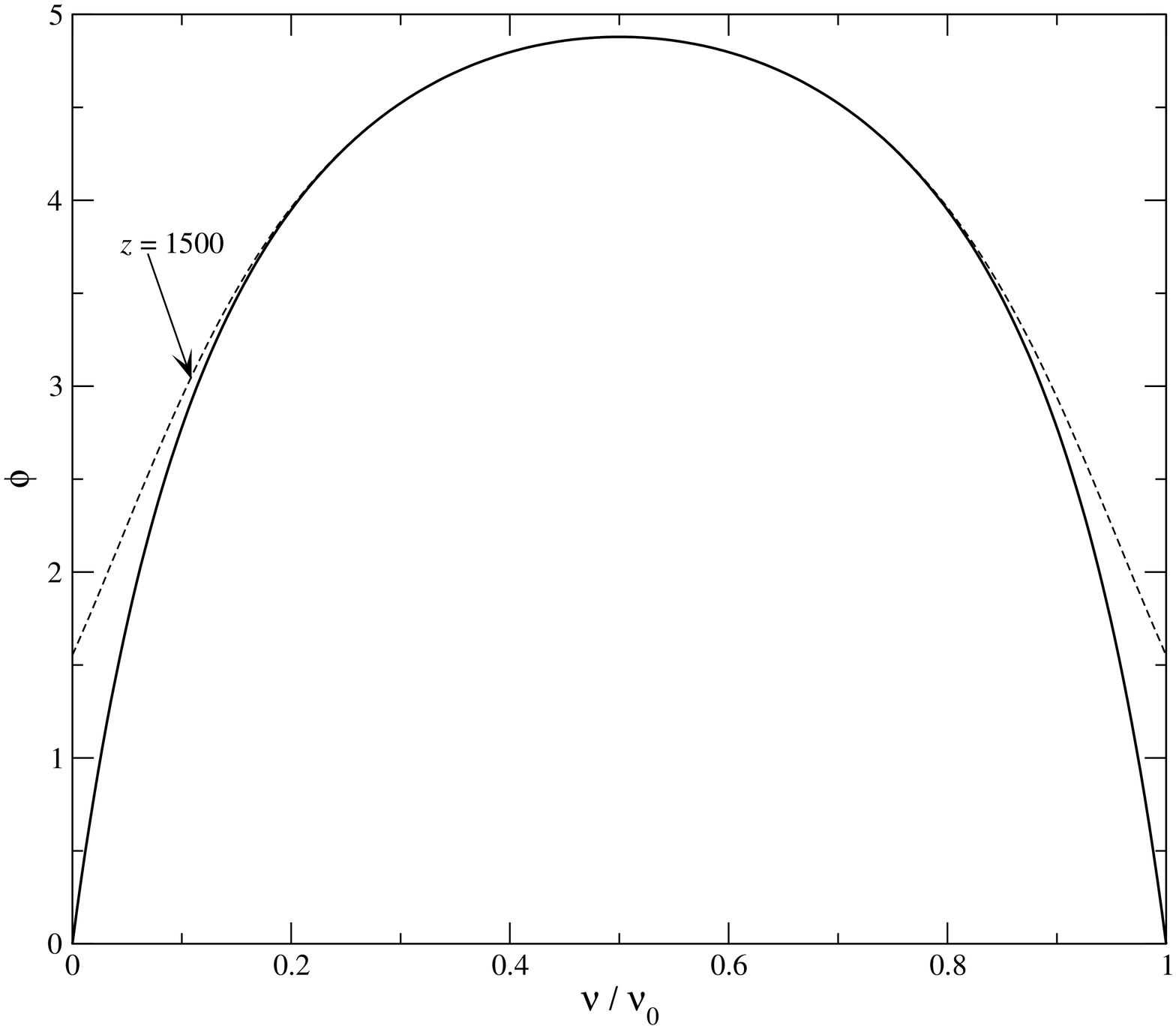}
  \caption{
    {\it Upper panel} -- $l$-dependence of the recombination coefficient,
    $\alpha_{nl}$, at $z=1300$ for different shells. The curves have been
    re-scaled by the {\it total} recombination coefficient, $\alpha_{n,\rm
      tot}=\sum_l \alpha_{nl}$, and multiplied by $l_{\rm max}=n-1$ such that
    the 'integral' over $\xi=l/l_{\rm max}$ becomes unity. Also the results
    obtained within the Kramers' approximation, i.e. $\alpha_{nl}^{\rm K}={\rm
      const}\times[2l+1]$, and without the inclusion of stimulated
    recombination for $n=100$ are presented.
    {\it Lower panel} -- Two-photon decay profile for the 2s-level in hydrogen: the
    solid line shows the broad two-photon continuum assuming that there is no
    ambient radiation field. In contrast, the dashed line includes the effects
    of induced emission due to the presence of CMB photons at $z=1500$. The
    figures are from \citet{RS_Chluba2006a} and \citet{RS_Chluba2007}.}
\label{RS:fig:Rci_Rci_tot}
\label{RS:fig:2s}
\end{figure}
%-----------------------------
In the transition of electrons from the 2s-level to the ground state two
photons are emitted in a broad continuum (see Fig. \ref{RS:fig:2s}). Due to
the presence of a large number of CMB photons at low frequencies, stimulated
emission becomes important when one of the photons is emitted close to the
Lyman-$\alpha$ transition frequency, and, as demonstrated in
\citet{RS_Chluba2006a}, leads to an increase in the effective two-photon
transition rate during hydrogen recombination by more than 1\%.
This speeds up the rate of hydrogen recombination, leading to a maximal change
in the ionization history of $\Delta N_{\rm e}/N_{\rm e}\sim -1.3\%$ at $z\sim
1050$.

\subsubsection{Re-absorption of Lyman-$\alpha$ photons.}
\label{RS:sec:Lya}
%---------------
The strongest distortion of the CMB blackbody spectrum is associated with the
Lyman-$\alpha$ transition and 2s-1s continuum emission. Due to redshifting
these access photons can affect energetically lower transitions.
%
%Although the re-absorption of photons in the Balmer-continuum is not important
%\citep{RS_Seager2000, RS_Chluba2007b}, 
%
The huge excess of photons in the Wien-tail of the CMB slightly increases the
${\rm 1s}\rightarrow{\rm 2s}$ two-photon absorption rate, resulting in
percent-level corrections to the ionization history during hydrogen
recombination \citep{RS_Kholu2006}.
%
%This feedback cancels a large part of the effect of induced two-photon decay
%of the 2s level as discussed in the previous paragraph.
%

\subsubsection{Feedback within the Lyman-series.}
\label{RS:sec:Lynfeedback}
%---------------
Due to redshifting, all the Lyman-series photons emitted in the transition of
electrons from levels with $n>2$ have to pass through the next lower-lying
Lyman-transition, leading to additional feedback corrections like in the case
of Lyman-$\alpha$ absorption in the 2s-1s two-photon continuum.
However, here the photons connected with Ly$n$ are completely absorbed by the
Ly$(n-1)$ resonance and eventually all Ly$n$ photons are converted into
Lyman-$\alpha$ or 2s-1s two-photon decay quanta.
As shown in \citet{RS_Chluba2007b}, this process leads to a maximal correction
to the ionization history of $\Delta N_{\rm e}/N_{\rm e}\sim 0.22\%$ at $z\sim
1050$.
%

%-----------------------------
\begin{figure}
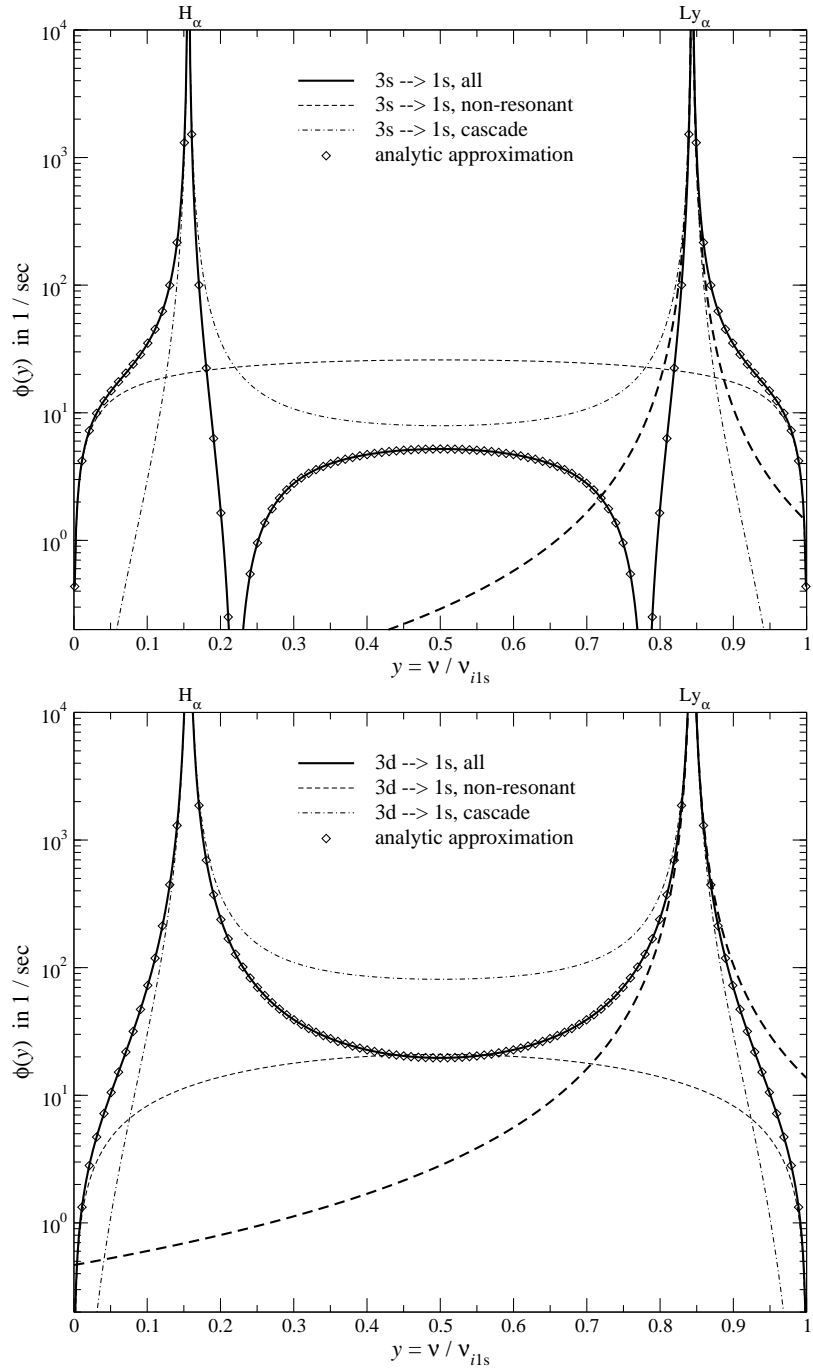

  \centering 
  \includegraphics[width=0.8\columnwidth]{./RS.3s.eps}
  \includegraphics[width=0.8\columnwidth]{./RS.3d.eps}
  \caption{Two-photon emission profile for the $3{\rm s}\rightarrow 1{\rm s}$ and $3{\rm
      d}\rightarrow 1{\rm s}$ transition. The non-resonant, cascade and
    combined spectra are shown as labeled. Also we give the analytic
    approximation as given in \citet{RS_Chluba2007c} and show the usual
    Lorentzian corresponding to the Lyman-$\alpha$ line (long dashed). 
%    
%    The resonances correspond to the Balmer-$\alpha$ and Lyman-$\alpha$
%    transition.  
%    
    The figure is from \citet{RS_Chluba2007c}.}
\label{RS:fig:3s3d}
\end{figure}
%-----------------------------
\subsubsection{Two-photon transitions from higher levels.}
\label{RS:sec:2ngamma2s}
%---------------
One of the most promising additional corrections to the ionization history is
due to the two-photon transition of highly excited hydrogen states to the
ground state as proposed by \citet{RS_Dubrovich2005}.
The estimated correction was anticipated to be as large as $\sim 5\%$ very
close to the maximum of the Thomson visibility function, and therefore should
have had a large impact on the theoretical predictions for the CMB power
spectra.
It is true that in the extremely low density plasmas the cascade of permitted
transitions (for example the chain 3s$\rightarrow$2p$\rightarrow$1s) goes
unperturbed and might be considered as two photon process with two resonances
\citep{RS_Goeppert}. In addition there is a continuum analogues to 2s-1s decay
spectrum and interference term between resonances and this weak continuum (see
Fig.  \ref{RS:fig:3s3d} and \citet{RS_Chluba2007c}).
However, the estimates of \citet{RS_Dubrovich2005} only included the
contribution to the two-photon decay rate coming from the two-photon continuum
due to virtual transitions, and as shown in \citet{RS_Chluba2007c} in
particular the interference between the resonance and non-resonance continuum
plays an important role in addition.
It is the departures of the two-photon line profile from the Lorentzian shape
in the very distant red wing of the Lyman-$\alpha$ line that matters, and the
overall corrections to the ionization history likely do not reach the
percent-level.
%%
%Interestingly, the two-photon decay of s-states typically leads to a decrease
%of the recombination rate, while d-states increase it. Overall the
%recombination rate increases due to the inclusion of this process, and in
%contrast to the conclusion of earlier considerations the effect has to be
%taken into account only for the first 5 shells.

%\subsubsection{Escape of Lyman-$\alpha$ photons.}
%\label{RS:sec:Lya-escape}
%%---------------

\section{Additional sources of uncertainty during hydrogen recombination.}
\label{RS:sec:uncertainty}
%---------------

\subsubsection{Feedback due to helium lines.}
\label{RS:sec:Hefeedback}
%---------------
The feedback of high frequency photons released during helium recombination
should also affect the dynamics of hydrogen recombination.
Here it is interesting that most of the photons from
\ion{He}{iii}~$\rightarrow$~\ion{He}{ii} will be already re-processed by
neutral helium before they can directly affect hydrogen.
However, those photons emitted by neutral helium can directly feedback on
hydrogen, but in order to take this feedback into account more detailed
computations of the helium recombination spectrum are required. 
This process should affect the hydrogen recombination history on a level
exceeding 0.1\%.

\subsubsection{Uncertainty in the CMB monopole temperature.}
\label{RS:sec:T0}
%---------------
The recombination history depends exponentially on the exact value of the CMB
monopole. As shown in \citet{RS_Chluba2007d} even for an error $\sim 1\,$mK
this leads to percent-level corrections in the ionization history.
However, the effect on the CMB power spectra is only slightly larger than
0.1\%.
At this level also the uncertainties in the helium abundance ratio, $Y_{\rm
  p}$, and the effective number of neutrinos, $N_\nu$, should also be
considered.  In particular the value of $Y_{\rm p}$ has a rather strong
impact, because it directly affects the peak of the Thomson visibility
function due to the dependence of the number of hydrogen nuclei, $N_{\rm H}$,
for a fixed number of baryons on the helium abundance ratio ($N_{\rm H}
\propto [1-Y_{\rm p}]$).

%
%In Sect. \ref{RS:sec:spectrum_cosmos} we already discussed the possibility to
%determine the value of the CMB monopole using the lines emitted during the
%recombination of hydrogen.
%%
%Similarly the primordial helium abundance could be measurable, but for this
%still the contributions from helium to the cosmological recombination spectrum
%should be studied.

\section{Conclusions.}
\label{RS:sec:conclusion}
%---------------
It took several decades until measurements of the CMB temperature fluctuations
became a reality.
After {\sc Cobe} the progress in experimental technology has accelerated by
orders of magnitude. 
Today CMB scientists are even able to measure $E$-mode polarization, and the
future will likely allow to access the $B$-mode component of the CMB in
addition.
Similarly, one may hope that the development of new technologies will render
the consequences of the discussed physical processes observable.
Therefore, also the photons emerging during the epoch of cosmological
(hydrogen) recombination could open another way to refine our understanding of
the Universe.
As we illustrated in this contribution, by observing the CMB spectral
distortions from the epoch of cosmological recombination we can in principle
directly measure cosmological parameters like the value of the CMB monopole
temperature, the specific entropy, and the pre-stellar helium abundance, {\it
  not suffering} from limitations set by {\it cosmic variance}.
Furthermore, we could directly test our detailed understanding of the
recombination process using {\it observational data}.
It is also remarkable that the discussed CMB signal is coming from redshifts
$z\sim 1300-1400$, i.e. before bulk of the CMB angular fluctuations were
actually formed.
To achieve this task, {\it no absolute measurement} is necessary, but one only
has to look for a modulated signal at the $\sim\mu$K level, with
typical amplitude of $\sim 30\,$nK and $\Delta \nu/\nu\sim 0.1$, where this
signal can be predicted with high accuracy, yielding a {\it spectral template}
for the full cosmological recombination spectrum, also including the
contributions from helium.

\acknowledgements %%% Text of acknowledgments runs on after this command.
The authors wish to thank Jos\'e Alberto Rubi\~no-Mart\'{\i}n for useful
discussions.
We are also grateful for discussion on experimental possibilities with
J.~E.~Carlstrom, M.~Pospieszalski and A.~Readhead during the NRAO
meeting.

%%% THE BIBLIOGRAPHY
%%%
%%% CONSULT SECTION 3 OF "INSTRUCTIONS FOR AUTHORS" FOR HOW TO USE NATBIB.
%%% AUTHORS ARE ENCOURAGED TO USE EITHER THE "THEBIBLIOGRAPY" ENVIRONMENT
%%% BY UNCOMMENTING (DELETING THE "%" SYMBOL) THE COMMANDS BELOW, OR BY
%%% USING THE BIBTEX ENVIRONMENT. TO FIND OUT WHICH IS APPLICABLE TO YOUR
%%% CONTRIBUTION, CONSULT THE VOLUME EDITORS FOR YOUR PROCEEDINGS.
%%%

\end{document}